\documentclass[%
   , colorlinks 
]{mpi2015-cscpreprint}

\usepackage[american]{babel}

\usepackage{graphicx}

\usepackage{amssymb}
\usepackage{amsthm}

\AtBeginDocument{

  \newcommand{\tlapack}{<T>LAPACK}
}

\usepackage{listings}
\usepackage{tikz}
\usetikzlibrary{positioning}
\usepackage{caption}
\usepackage{subcaption}

\begin{document}

\title{Low Precision Fortran -- Enabling Low Precision Floating Point
Arithmetic in Modern Fortran}

\author[$\ast$]{Martin K\"ohler}
\affil[$\ast$]{Max Planck Institute for Dynamics of Complex Technical Systems,
    Magdeburg, Germany,\authorcr
    \email{koehlerm@mpi-magdeburg.mpg.de}, \orcid{0000-0003-2338-9904}}
\author[$\dagger$]{Peter Benner}
\affil[$\dagger$]{Max Planck Institute for Dynamics of Complex Technical Systems,
    Magdeburg, Germany,\authorcr
    \email{benner@mpi-magdeburg.mpg.de}, \orcid{0000-0003-3362-4103}}

\shorttitle{Low Precision Fortran}
\shortauthor{M. Köhler, P. Benner}
\shortdate{}


\abstract{%
  Although Fortran is almost 70 years old, the language continues to evolve
  in order to keep pace with developments in computer science. In particular, a
  flexible
  type system was introduced that allows developers to specify the sizes of
  floating-point numbers and integers. In the latest revisions of the Fortran
  standard\cite{Fortran2023}, portable type variants for IEEE
  754~\cite{IEEE754-2019} binary64 (double precision, \texttt{real64}) and
  binary32 (single precision, \texttt{real32}) were added. However, the rapid
  development of AI toolkits and accelerator hardware has created a strong
  focus on floating-point types of lower precision and lower memory usage than
  binary32.

  While the IEEE 754-2019~\cite{IEEE754-2019} standard defines the binary16
  type
  for representing
  half-precision numbers, the Fortran standard does not provide the
  \texttt{real16} variant in the type system. In contrast, most C compilers
  support such a data
  type. In numerical linear algebra, there is strong interest in exploiting
  the high performance of accelerator devices for core algorithms like matrix
  decompositions or iterative solvers. Especially when the performance
  ratio between double, single, and half precision is on the order of 1:2:20,
  as on current NVidia H100 accelerators,
  it becomes highly beneficial to use lower-precision types. Yet, before
  performance can be targeted, correctness and accuracy must be verified when
  operating below single precision.

  In an ideal world, existing codes from BLAS, LAPACK~\cite{Lapack}, and
  similar
  libraries could simply be adapted by changing data types and function names
  (or prefixes). In practice, however, this often requires rewriting
  significant portions of code to fit into accelerator software frameworks
  -- a learning-intensive and time-consuming process. This is not a satisfying
  approach when developing proofs of concept or running exploratory
  experiments that reuse existing code.

  In this article, we present our Low Precision Fortran (LPF)
  library~\cite{lpf} that enables the use of low-precision types --
  \texttt{binary16}, \texttt{bfloat16}, \texttt{fp8\_e4m3}, and
  \texttt{fp8\_e5m2} -- just like any other floating-point type in Fortran.
  Furthermore, we introduce extensions that support BLAS operations in low
  precision and show how easily existing routines can be rewritten to use these
  data types.
}

\novelty{The Low Precision Fortran library extends the Fortran ecosystem by
providing a seamless integration of low precision floating point types. In our
contribution we cover:
\begin{itemize}
 \item The integration of IEEE-754 binary16, BFloat16, FP8\_E4M3, and FP8\_E5M3
     into Fortran just like any other floating point number.
 \item The realization of the Basic Linear Algebra Subroutines (BLAS) on top of
     the low precision types.
 \item An interface layer such that a seamless interoperability with C++ is
     possible.
 \item The focus for all types is to use representations that are (or can be)
     implemented in hardware.
\end{itemize}}

\maketitle

\section{Motivation}
Although Fortran exists for nearly 70 years, the programming language
continues to evolve and adapt to the needs of modern software engineering. The
language and its type system support various floating-point types by specifying
the desired type as a ``kind'' parameter. In modern Fortran, since the
Fortran~90 release, this is achieved via:
\begin{lstlisting}[language=fortran]
real(kind = X) :: foo
\end{lstlisting}
where \texttt{X} specifies the floating-point type. To ensure portability,
Fortran 2008~\cite{Fortran2008} introduced the \texttt{iso\_fortran\_env}
module, which provides portable ``kind'' definitions for various data types.
Specifically for floating-point types, it defines \texttt{real32},
\texttt{real64}, and \texttt{real128}, corresponding to the IEEE~754
\emph{binary32}, \emph{binary64}, and \emph{binary128} formats, respectively.
Recent hardware developments and the increasing focus on AI applications have
made low-precision floating-point formats (i.e., formats using less than four
bytes) increasingly prevalent. Unfortunately, this pace of development often
exceeds the standardization process of most programming languages;
consequently, even the Fortran 2023~\cite{Fortran2023} standard does not
include
a \texttt{real16} kind or similar definitions to provide an interface to the
IEEE 754 half-precision type~\cite{IEEE754-2019}, Google's
BFloat16~\cite{WanK19}, or the quarter-precision FP8 formats of the Open
Compute Project~\cite{MicOD23,MicSBetal22}. To the authors' knowledge, as of
2026, only the NAG Fortran compiler provides a non-standard extension to
support IEEE 754 half-precision, but no other low-precision types.

Recent hardware advancements demonstrate that half- and quarter-precision
formats significantly outperform single- and double-precision data types in
terms of runtime and memory consumption. The
traditional assumption that halving the size of a floating-point type would
roughly double performance no longer holds. For every double-precision
floating-point operation (FLOP) performed on GPU accelerators -- such as the
NVIDIA H100
GPU\footnote{\url{https://resources.nvidia.com/en-us-gpu-resources/h100-datasheet-24306},
accessed 05-05-2026} --  one can perform two single precision, 55
half-precision,
or 111 quarter-precision FLOPs. Similar gains are evident on general-purpose
CPUs; for instance, Intel's 4th generation Xeon
CPUs\footnote{\url{https://www.intel.com/content/www/us/en/developer/articles/technical/use-new-built-in-ai-acceleration-engines.html}}
 can perform up to nine BFloat16 FLOPs for every double-precision FLOP.

This shift presents a significant challenge: the need to develop new algorithms
or adapt existing ones in scientific computing to leverage this evolving
computational landscape while maintaining the required numerical accuracy. The
\emph{Low Precision Fortran}~\cite{lpf} library aims to provide an intuitive
interface for low-precision floating-point types in Fortran that integrates
seamlessly into existing infrastructure. Our focus is on enabling the portable
evaluation and development of algorithms, ensuring that the behavior of a
specific data type is emulated even in the absence of native hardware support.
Consequently, the effects of low-precision types can be examined without
requiring specialized hardware or enduring the steep learning curve of
hardware-agnostic toolkits.

A primary design objective of the library is to keep these data types as
compatible as possible with standard \texttt{real} types. In other words, once
the Fortran standard supports a specific low-precision type, the necessary
changes to existing code should be minimal ($\mathcal{O}(1)$). Ideally, only
the variable declarations need to be updated, leaving the remainder of the code
untouched. In the following sections, we describe the implementation of this
approach and provide examples demonstrating how easily algorithms can be
formulated using low-precision data types.

In Section~\ref{sec:user-defined-types}, we explain how the type system of
modern Fortran can be used and list the requirements that need to be fulfilled
to make user defined data types behaving like a floating point type. The third
section covers two different strategies to implement the required operations on
these data types with focus on interoperability and potential hardware support.
Section~\ref{sec:blas}, we show how the BLAS functionality can be
reimplemented on top of the user defined data-types and how high-level linear
algebra operations  can be imported from \tlapack. Finally, a usage example in
Section~\ref{sec:usage} shows low-precision data types can be used as usual
with our library.

\section{User-Defined Data Types in Fortran}
\label{sec:user-defined-types}
When thinking of Fortran, many recall the legacy style of Fortran 77, which
supported only basic scalar and array types. Over the last three decades, this
perception has evolved. Beginning with Fortran 90, user-defined data types were
introduced via the \texttt{type} statement. Subsequent standards introduced
function and operator overloading, as well as the ability to define custom
handlers to support \texttt{read}, \texttt{print}, and \texttt{write}
statements for user-defined types. Function and operator overloading can
be implemented using the \texttt{interface} construct. Essentially, this
mechanism functions similarly to the multiple-dispatch approach in the Julia
programming language~\cite{Julia-2017} or operator overloading in C++. An
interface defines a ``top-level'' name associated with a set of functions. When
an interface is invoked as a function or subroutine, the Fortran compiler
matches the types of the arguments against the available function and
subroutine definitions. Consequently, functions such as $\sin$ and $\cos$, as
well as operators like $+$ and $-$, can be implemented for new data types
without altering the behavior of existing implementations. However, this
approach requires the implementation of all meaningful combinations of function
arguments. In this context, ``meaningful'' refers both to mathematical
necessity and to the programmer's desire for concise expression. For example,
consider the addition of two floating-point numbers, $a$ and $b$, where at
least one is a user-defined type. The resulting combinations for implementing
the addition operator are shown in Table~\ref{tab:add-combi}. Given that IEEE
binary128~\cite{IEEE754-2019} (quad precision) is rarely used and provides
precision far beyond that of lower-precision data types, combinations
involving \texttt{real128} are omitted.
\begin{table}[t]\centering
    \begin{tabular}{|c|c||c|c|}\hline
        $a$ & $b$ & $a$ & $b$ \\\hline
        user-defined & user-defined & user-defined & real32 \\
        user-defined & real64 & user-defined & real128 \\
        real32 & user-defined &  real64 & user-defined \\
        real128 & user-defined & user-defined & integer \\
        integer & user-defined & & \\\hline
    \end{tabular}
    \caption{data type combinations for $a+b$}
    \label{tab:add-combi}
\end{table}
Using the $+$ operator as an example, the following interface definition is
obtained:
\begin{lstlisting}[language=fortran]
interface operator(+)
  module procedure add_X_X
  module procedure add_X_real32, add_real32_X
  module procedure add_X_real64, add_real64_X
  module procedure add_X_int, add_int_X
end interface operator(+)
\end{lstlisting}
Here, \texttt{X} serves as a placeholder for the data type name. If
combinations with other, possibly user-defined, data types are required, they
can be added to this interface or be defined elsewhere. The pattern matching
used
to select the appropriate implementation for an operator is not restricted to a
single \texttt{interface} block. Without loss of generality, each
\texttt{module procedure} for $c=a+b$ must then be implemented as follows:
\begin{lstlisting}[language=fortran]
elemental function add_X_real32(a, b) result(c)
  type(X), intent(in) :: a
  real(real32), intent(in) :: b
  type(X) :: c
  ! Perform the add with a and b, writing the result to c
end function
\end{lstlisting}
Consequently, it is important that the function is defined as
\texttt{elemental}. On the one hand, this ensures that the function is free of
side effects; for example, it neither modifies global variables nor depends on
an internal, potentially changing, state. On the other hand, it allows the
function to be applied element-wise. This is necessary to support vector/array
expressions in Fortran, enabling the use of the data type in expressions such
as:
\begin{lstlisting}[language=fortran]
c(1:n) = a(1:n) + b(1:n)
\end{lstlisting}
Another important property of the operational definition described above is
that for expressions such as \texttt{b = a + 1.0} or \texttt{b = a +
1.0D0}~--~where \texttt{a} is of a low-precision data type and \texttt{1.0}
(\texttt{real32}) or \texttt{1.0D0} (\texttt{real64}) are constants~--~the
result \texttt{b} will also be of the low-precision data type. This prevents
precision leakage and allows for the direct use of literals in expressions
without the need to first store them in a low-precision variable. If higher
precision is required for the result or during computation, the user must
explicitly implement a typecast to the higher precision. To make the data type
as versatile as any other scalar value, the following operations should be
defined using the aforementioned scheme: $a+b$, $a-b$, $a\cdot b$, $a / b$,
$-a$, $a < b$, $a\le b$, $a > b$, $a \ge b$, $a \neq b$, $a == b$, and $a^b$.

To enable assignments, the assignment operator \text{=} must also be defined. A
key property of this operator is that it considers both sides of the
\texttt{=} operator, allowing for the implementation of the following two use
cases:
\begin{enumerate}
    \item Assigning a \texttt{real32}, \texttt{real64}, or \texttt{integer}
    value to a low-precision type. This results in an implicit typecast; for
    example, in the expression \texttt{x = 1.0D0}, the value \texttt{1.0D0} is
    automatically converted to the low-precision type.
    \item Assigning a low-precision variable to a \texttt{real32} or
    \texttt{real64} variable, which performs an implicit up-cast.
\end{enumerate}
The assignment operator is implemented similarly to the arithmetic operators,
and all such implementations must include the \texttt{elemental} attribute.
This enables the user to convert, for example, a matrix from \texttt{real32} to
a low-precision type as follows:
\begin{lstlisting}[language=fortran]
type(X), dimension(m,n) :: a_lp
real(real32), dimension(m,n) :: a
a_lp = a
\end{lstlisting}

Since values must be read from or written to files or the terminal, the I/O
operations for the \texttt{read}, \texttt{write}, and \texttt{print} statements
must also be provided. Fortran distinguishes between formatted and
unformatted I/O; the former is human-readable, while the latter is
machine-readable. Because the structure of unformatted files is not
standardized, the resulting file structure depends on the compiler.
Consequently, storing data in this manner does not align with the FAIR
principles of research data management~\cite{BarCKetal22}. For this reason,
unformatted I/O was not implemented by now. Regarding the human-readable
interface,
two subroutines must be defined:
\begin{lstlisting}[language=fortran]
interface write(formatted)
  module procedure write_formatted
end interface
interface read(formatted)
  module procedure read_formatted
end interface

subroutine write_formatted(dtv,unit,iotype,vlist,iostat,iomsg)
subroutine read_formatted(dtv,unit,iotype,vlist,iostat,iomsg)
\end{lstlisting}
Here, \texttt{dtv} represents the value to be read or written, \texttt{unit}
specifies the file unit to be used, and \texttt{iotype} and \texttt{vlist}
control the formatting and the method by which the value is read or written.
Finally, the \texttt{iostat} and \texttt{iomsg} parameters are used for error
handling. To reduce complexity and leverage the existing formatting
functionality of the Fortran standard library, these routines operate
internally with \texttt{real32}. This approach allows for the handling of
low-precision types as follows:
\begin{lstlisting}[language=fortran]
type(X) :: a
write(*,*) a
write(*,'(A, DT"F"(6,3))')  "Output with DTF(6,3) A =", a
write(*,'(A, DT"E"(6,3))')  "Output with DTE(6,3) A =", a
\end{lstlisting}

The final remaining components are the intrinsic functions. These include
mathematical functions such as $\sin$ and $\cos$, functions that return
numerical properties of the data type (e.g., \texttt{epsilon} and
\texttt{huge}), and reduction functions such as \texttt{minloc} and
\texttt{minval}. By defining an interface as shown below, these routines are
made available for the user-defined data type:
\begin{lstlisting}[language=fortran]
interface sin
  module procedure sin_X
end interface
elemental function sin_x( a )  result(out)
  type(X), intent(in) :: a
  type(X) :: out
  !... sin implementation
end function
\end{lstlisting}
In most cases, these functions must be defined as \texttt{elemental}.
Functions that handle array-valued inputs only need to be \texttt{pure} to
ensure they are free of side effects. Depending on the nature of the function,
they are implemented using one of the following approaches:
\begin{enumerate}
    \item The functionality is implemented using basic operations. This is the
    case for \texttt{abs}, \texttt{min}, \texttt{minloc}, and similar
    functions.
    \item The function returns properties of the data type, which are provided
    as hard-coded values. This applies to functions such as \texttt{epsilon}
    and \texttt{huge}.
    \item The function accesses the bit-wise representation of the data type.
    This method is used to implement \texttt{fraction}, \texttt{exponent}, and
    others.
    \item The value is retrieved from a lookup table or computed via a
    round-trip conversion through single precision:
    \[
    \text{low-precision} \rightarrow \text{single precision} \rightarrow
    \text{function evaluation} \rightarrow \text{low-precision}
    \]
    All mathematical functions are implemented in this manner. Depending on the
    size of the data type, a lookup table is used for 1-byte types, while the
    round-trip method is employed for all larger types.
\end{enumerate}
To ensure the user-defined data type serves as a full replacement, the
following functions must be implemented as described above:
\texttt{abs}, \texttt{acos}, \texttt{acosh}, \texttt{asin}, \texttt{asinh},
\texttt{atan}, \texttt{atanh}, \texttt{bessel\_j0}, \texttt{bessel\_j1},
\texttt{bessel\_y0}, \texttt{bessel\_y1}, \texttt{ceiling}, \texttt{cos},
\texttt{cosh}, \texttt{erf}, \texttt{erfc}, \texttt{exp}, \texttt{floor},
\texttt{gamma}, \texttt{log}, \texttt{log10}, \texttt{log\_gamma},
\texttt{sin},
\texttt{sinh}, \texttt{sqrt}, \texttt{tan}, \texttt{tanh}, \texttt{atan2},
\texttt{bessel\_jn}, \texttt{bessel\_yn}, \texttt{hypot}, \texttt{radix},
\texttt{epsilon}, \texttt{exponent}, \texttt{fraction}, \texttt{cosd},
\texttt{sind}, \texttt{tand}, \texttt{cotan}, \texttt{cotand},
\texttt{acosd}, \texttt{asind}, \texttt{atand}, \texttt{atan2d},
\texttt{erfc\_scaled}, \texttt{huge}, \texttt{tiny}, \texttt{digits},
\texttt{minexponent}, \texttt{maxexponent}, \texttt{mod}, \texttt{modulo},
\texttt{nearest}, \texttt{nint}, \texttt{precision}, \texttt{range},
\texttt{scale}, \texttt{sign}, \texttt{maxval}, \texttt{maxloc},
\texttt{minval}, \texttt{minloc}, \texttt{isnan}, \texttt{isinf},
\texttt{min}, \texttt{max}, \texttt{matmul}, \texttt{norm2},
\texttt{dot\_product}, \texttt{product}, and \texttt{sum}.

Finally, there are a few aspects of the implementation that are not entirely
seamless. First, Fortran does not support custom suffixes for scalar values to
specify how they should be interpreted. For example, while one can write
\texttt{1.0\_real64} to guarantee that a value is interpreted as double
precision, using a user-defined suffix such as \texttt{1.0\_X} is not
supported. This may cause difficulties in pattern matching when selecting the
correct implementation from a generic interface. For instance, the following
code:
\begin{lstlisting}[language=fortran]
type(X) :: a
a = 1.0
write(*,*) hypot(a, 2.0)
\end{lstlisting}
will prompt the compiler to search for a function where the first parameter is
of type \texttt{X} and the second is a single-precision \texttt{real32} value.
To simplify this process, one can define a function with the same name as the
data type that accepts integer, single, or double-precision inputs and returns
a value converted to the user-defined type. This serves as a constructor for
the type, similar to the approach used in languages such as C++.

Another issue concerns the conversion function \texttt{real(y, kind)}, where
the return type depends on the value of the \texttt{kind} parameter. This could
not be implemented because all types in the function signature must be
well-defined and independent of the input values. Consequently, we provide an
implementation of \texttt{real(y)} (without the \texttt{kind} parameter) to
convert a user-defined type to \texttt{real32}, and an implementation of
\texttt{dble(y)} to convert the type into a \texttt{real64} floating-point
value.

One language construct lacks proper support in this implementation of
user-defined floating-point types is the \texttt{parameter} property. Because
type conversions are evaluated at runtime, they cannot be utilized at compile
time. Consequently, the only way to define a constant is as follows:
\begin{lstlisting}[language=fortran]
type(X), parameter :: const_value = X( value = IntegerRepresentation )
\end{lstlisting}
Here, \texttt{IntegerRepresentation} is the integer value in memory that
represents the desired floating-point number.

Overall, this approach allows for the implementation of floating-point types
other than \texttt{real32}, \texttt{real64}, and \texttt{real128} that can be
used normally within the code. This method is not restricted to low-precision
types and can likewise be used to provide very high-precision types. To achieve
the goal of $\mathcal{O}(1)$ changes, the following elements must be refactored
once the Fortran compiler supports this functionality:
\begin{itemize}
    \item Change \texttt{type(X) :: \ldots} to \texttt{real(kind=X) :: \ldots}
    \item Remove the constructor function or provide a dummy function that
    returns its input cast to \texttt{real(kind=X)}.
    \item Adjust the format specifiers in \texttt{read}, \texttt{write}, and
    \texttt{print}.
\end{itemize}
These modifications can be performed automatically using common programming
tools and regular expressions. Consequently, the previously stated goal is
fulfilled, and the framework is ready to be populated with implementations of
specific low-precision types, such as \texttt{fp16}, \texttt{bf16}, and
\texttt{fp8}. The details of this implementation are provided in the next
section.

\section{Implementing Low-Precision Arithmetic}
\label{sec:implementation}
The floating-point arithmetic components of the interfaces can be implemented
using two different approaches. The first leverages the fact that C compilers
typically support a wider range of floating-point types than their
corresponding Fortran compilers (e.g., \mbox{gcc $\leftrightarrow$ gfortran},
\mbox{clang $\leftrightarrow$ flang}, or \mbox{icx $\leftrightarrow$ ifx}).
In
this approach,
arithmetic operations are implemented as concise C functions, allowing the
compiler to handle the underlying type precision. Alternatively, if such
support is unavailable, the operations are implemented either via specialized
libraries or by manipulating the bit-level representation of the data type.

For the first case, it is necessary, and for the second, it is useful, to
ensure seamless data interoperability between C and Fortran. Therefore, we
employ the \texttt{iso\_c\_binding}~\cite{Fortran2008,Fortran2023} module of
Fortran to establish a portable and stable interface between the two languages.
By using the \texttt{bind(C)} keyword, Fortran objects~--~such as types,
enumerators, and function declarations~--~are made compatible with C.
Furthermore,
the bit-width of our low-precision type must be 8, 16, 32, 64, or 128 bits. If
these constraints are not met, the leading bits must be zero-filled. For each
of these sizes, a corresponding integer type with the same bit-width exists in
C. This allows us to represent a low-precision value using the following type
in Fortran:
\begin{lstlisting}[language=fortran]
type, bind(c) :: fpX
  integer(kind = c_intX_t) :: value
end type
\end{lstlisting}
where \texttt{X} represents the aforementioned bit-widths. It is important to
note that in Fortran, unless specified otherwise, arguments are passed by
reference (as pointers) in function calls. For the type \texttt{fpX}, calling a
C function from Fortran:
\begin{lstlisting}[language=fortran]
type(fpX) :: a
call some_c_function(a)
\end{lstlisting}
corresponds to the following C definition:
\begin{lstlisting}[language=c]
void some_c_function(struct fpX *a);
\end{lstlisting}
where \lstinline[language=c]!struct fpX {intX_t value;}!. Because the structure
contains only a single element, the following properties hold:
\begin{enumerate}
    \item The structure is not padded; i.e.,
    \lstinline[language=c]!sizeof(intX_t)==sizeof(struct fpX)!.
    \item A pointer to the structure is equivalent to a pointer to an integer
    of the same size.
\end{enumerate}
Consequently, the C function can also be written as:
\begin{lstlisting}[language=c]
    void some_c_function(intX_t *a);
\end{lstlisting}
which simplifies access to \texttt{a}. Furthermore, this approach facilitates
the easy transfer of not only scalar values but also arrays between Fortran and
C. Specifically, a vector defined in Fortran AS ASSUMED-SIZE is represented
as a pointer to its
first element in C. For an array of type \texttt{fpX}, the following mapping
applies:
\begin{equation}
    \label{eqn:vector}
    \underbrace{\text{vector\_a}(k)}_{\text{Fortran}}  \equiv
    \underbrace{\text{vector\_a}[k-1]}_{\text{C}}.
\end{equation}
Because Fortran treats higher-dimensional objects as one-dimensional arrays
using a leading-dimension-based index mapping, these map to C in the same
manner. This simplifies the subsequent implementation of BLAS operations and
ensures that our low-precision data types follow the same memory layout as
existing floating-point types. Since this approach focuses on the bit-level
representation of data types to enable direct hardware mapping, it differs from
libraries such as GNU MPFR~\cite{FouHLetal07}, which employ more complex
structures. Table~\ref{tab:bit-types} lists the types implemented in Low
Precision Fortran and their corresponding counterparts in C.
\begin{table}\centering
    \begin{tabular}{|l|c|c|c|c|}\hline
        & size in bits & Fortran integer kind & C integer type & C float
        type\\\hline
    IEEE binary16~\cite{IEEE754-2019} & 16 & \texttt{c\_int16\_t} &
    \texttt{int16\_t} & \texttt{\_Float16} \\
    BFloat16~\cite{WanK19} & 16 & \texttt{c\_int16\_t} & \texttt{int16\_t} &
    \texttt{\_\_bf16} \\
    OCP FP8\_E4M3~\cite{MicOD23} & 8 & \texttt{c\_int8\_t} &
    \texttt{int8\_t} &
    -- \\
    OCP FP8\_E5M2~\cite{MicOD23} & 8 & \texttt{c\_int8\_t} &
    \texttt{int8\_t}
    & -- \\\hline
    \end{tabular}
    \caption{Low-precision types and the Fortran/C representation}
    \label{tab:bit-types}
\end{table}

\subsection{Implementing Types with C Compiler Support}
As previously mentioned, several floating-point types are natively supported
by
C compilers. Referring to Table~\ref{tab:bit-types}, this includes IEEE
binary16 (fp16) and Google's BFloat16 (bf16). Depending on hardware support,
the compiler either generates the corresponding machine instructions or
provides an emulation layer. In both cases, standard arithmetic operations (+,
-, *, /) and type conversions remain applicable. Without loss of generality,
the workflow described herein focuses specifically on the IEEE binary16 type.

To avoid indirect memory access caused by passing pointers to operands, we use
the \texttt{value} property in the Fortran interfaces to enforce pass-by-value
instead of pass-by-reference. A naive approach would be to implement a wrapper
function in C:
\begin{lstlisting}[language=c]
_Float16 add_fp16_fp16(_Float16 a, _Float16 b) {
  return a + b;
}
\end{lstlisting}
with the following \texttt{bind(C)} interface on the Fortran side:
\begin{lstlisting}[language=fortran, morekeywords={bind}]
interface
  function help_add_fp16_fp16(a, b) result(c) &
    & bind(C,name='add_fp16_fp16')
    integer(kind = c_int16_t), intent(in), value :: a, b
    integer(kind = c_int16_t) :: c
  end function
 end interface

function add_fp16_fp16(a, b) result(c)
  type(fp16), intent(in) :: a, b
  type(fp16) :: c
  c%value = help_add_fp16_fp16(a%value, b%value)
end function
\end{lstlisting}
Unfortunately, this fails on at least the x86\_64 architecture, even though the
sizes of \texttt{\_Float16} and \texttt{int16\_t} are identical. This is due to
the System V Application Binary Interface (ABI)~\cite{SysV-ABI} for
machine-level function argument passing, which is adopted by many operating
systems. The ABI specifies that the first two integer parameters are passed in
CPU registers \texttt{rdi} and \texttt{rsi}, whereas the first two
floating-point arguments are expected to reside in \texttt{xmm0} and
\texttt{xmm1}. Consequently, the data passed from Fortran is placed in a
register that the C function does not read. Even Microsoft Windows, which does
not inherently follow the System V ABI, employs a similar mechanism.

Although some platforms may be exempt from this issue, a portable and unified
solution is required to overcome this problem without introducing additional
overhead. A possible solution is to leverage the memory-level equivalence of
\texttt{\_Float16} and \texttt{int16\_t}. Both types require two bytes and can
be treated as a sequence of 16 bits. This leads to the use of a C
\lstinline[language=c]!union! to map the floating-point type to the
corresponding integer type:
\begin{lstlisting}[language=c]
typedef union {
  int16_t i16;
  _Float16 f16;
} fp16_handler_t;
\end{lstlisting}
Our helper function in C is then defined as:
\begin{lstlisting}[language=c]
int16_t add_fp16_fp16(int16_t _a, int16_t _b) {
  fp16_handler_t a = {.i16 = _a}, b = {.i16 = _b}, c;
  c.f16 = a.f16 + b.f16;
  return c.i16;
}
\end{lstlisting}
This approach is invariant to the parameter-passing conventions of the ABI.
This is not the case for array-valued functions, as arrays are passed by
reference by definition, and pointers are stored identically regardless of the
type they reference.

In addition to standard arithmetic operations, transcendental functions such as
$\sin(X)$ are also required. While many modern CPUs support basic arithmetic
for
these types, extended functionality is generally not available. Even standard
math libraries typically only provide functions for IEEE binary32, binary64,
and occasionally binary128. To implement these functions, one could either
derive them from mathematical definitions using basic operations, utilize
lookup tables for each function value, or employ a round-trip conversion via
the binary32/float32 type. The first approach is time-consuming and error-prone
and is therefore not adopted here. The second approach is feasible for
functions with a single argument; since each binary16 value can be mapped to an
integer, a lookup table would contain 65,536 entries, requiring 128 KiB of
memory. However, given the number of functions mentioned in
Section~\ref{sec:user-defined-types}, the total memory requirement would reach
several megabytes. Furthermore, for functions with two binary16 inputs, the
resulting lookup table would require 2 GiB, which is impractical. Consequently,
for two-byte low-precision types, we use the third approach: implementing a
round-trip via the binary32 data type. For the $\sin$ function, this leads to
the following helper function in C:
\begin{lstlisting}[language=c]
int16_t sin_fp16(int16_t _x) {
  fp16_handler_t a = {.i16 = _x}, sinx;
  sinx.f16 = (_Float16) sinf((float) a.f16);
  return sinx.i16;
}
\end{lstlisting}
Since support for the \texttt{\_Float16} type is provided by the C compiler,
the typecasts are handled by the compiler as well. The Fortran part of the
implementation
is analogous to the \text{add\_fp16\_fp16} function described above. Following
the same semantics, functions for the conversion between binary32 and binary16
are also implemented.

\paragraph{Remark on Compiler and Hardware Support}
Most modern C compilers, including the GNU Compiler Collection (GCC),
LLVM/Clang, and the Intel C Compiler, provide support for the `binary16` and
`BFloat16` data types. On x86\_64 hardware, this support requires the
`\texttt{-mf16c}` compiler flag and a CPU compatible with the x86-64-v2
architecture
specifications for proper emulation. If the CPU natively supports `binary16`
or `BFloat16` arithmetic, additional compiler flags must be specified. For
instance, when using GNU compilers on CPUs with AVX512-FP16 or AVX512-BF16
support, the
`\texttt{-mavx512fp16}` and `\texttt{-mavx512bf16}` flags can be employed. For
Aarch64 CPUs, such as Apple Silicon, the
`\texttt{-march=armv8.4-a+fp16+bf16}` flag is used. Regardless of whether the
hardware natively supports these low-precision types, the
`\texttt{-fexcess-precision=16}` flag should be used to ensure the compiler
does not employ higher precision for intermediate results during `binary16` or
`BFloat16` computations.
When targeting other hardware or compilers, refer to the respective technical
documentation for the appropriate compiler options.

\subsection{Implementing Types without C Compiler Support}
\label{subsec:nocompiler}
For lower-precision types such as FP8\_E4M3 or FP8\_E5M2~\cite{MicSBetal22},
compiler support is typically non-existent. Therefore, arithmetic
operations must be implemented directly on the bit representations. Both FP8
types offer saturating and non-saturating variants. While the former maps
infinity to the maximum representable value, the latter handles infinity
conventionally. Because the saturating mode is specific to AI applications, we
use
the non-saturating mode for consistency with standard floating-point behavior.
Without loss of generality, we describe the FP8\_E5M2 type here, following the
Open Compute Project guidelines~\cite{MicSBetal22}.

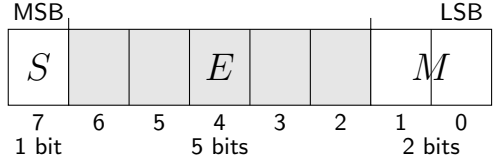
\begin{figure}
    \centering
    \begin{tikzpicture}[
    font=\sffamily\small
]
\draw[fill=white]   (0,0) rectangle ++(0.8,1);   
\draw[fill=gray!20] (0.8,0) rectangle ++(4,1);   
\draw[fill=white]   (4.8,0) rectangle ++(1.6,1); 

\foreach \x in {1.6,2.4,3.2,4.0,5.6}
    \draw (\x,0) -- (\x,1);

\node at (0.4,0.5) {\Large $S$};
\node at (2.8,0.5) {\Large $E$};
\node at (5.6,0.5) {\Large $M$};

\node[above] at (0.4,1) {\small MSB};
\node[above] at (6.0,1) {\small LSB};

\foreach \x/\b in {
    0.4/7,
    1.2/6,
    2.0/5,
    2.8/4,
    3.6/3,
    4.4/2,
    5.2/1,
    6.0/0}
{
    \node[below] at (\x,0) {\small \b};
}

\draw (0.8,1) -- ++(0,0.15);
\draw (4.8,1) -- ++(0,0.15);

\node[below,yshift=-2ex] at (0.4,0) {\small 1 bit};
\node[below,yshift=-2ex] at (2.8,0) {\small 5 bits};
\node[below,yshift=-2ex] at (5.6,0) {\small 2 bits};
\end{tikzpicture}
    \vskip-5ex
    \caption{Storage mapping of the components of an FP8\_E5M2 number in an
    8-bit integer.}
    \label{fig:fp8e5m2}
\end{figure}

For all computations, we adopt the following representation for the
floating-point number $x$:
\begin{equation}
    \label{eqn:fp-repr}
    x = \left(-1\right)^S \cdot 2^{E-B} \cdot\left(1+2^{-m} \cdot M\right).
\end{equation}
If the number is subnormal (i.e., $E=0$), $x$ is expressed as:
\begin{equation}
    x = \left(-1\right)^S \cdot 2^{1-B} \cdot\left(1+2^{-m} \cdot M\right).
\end{equation}
The parameters used in these equations are defined as follows:
\begin{itemize}
    \item $S$ is a single bit indicating the sign.
    \item $E$ is the stored exponent, represented as a five-bit unsigned
    integer, where $0\leq E\leq 31$.
    \item $B$ is the exponent bias. Since $E$ is stored as an unsigned integer,
    a bias is used to avoid two's-complement integer arithmetic. The actual
    exponent is given by $E-B$, as shown in Equation~\eqref{eqn:fp-repr}.
    \item $M$ is the mantissa, stored as a two-bit unsigned integer.
    \item $m$ denotes the number of bits in the mantissa.
\end{itemize}
Since $m$ and $B$ are constant for this format, only $S$, $E$, and $M$ need to
be stored. Figure~\ref{fig:fp8e5m2} illustrates how these components are mapped
into an eight-bit integer. These values can be easily accessed using bit-wise
operators in C.
In addition to the values of $E$ and $M$ that define normal numbers, several
special values exist to represent infinity ($\infty$), NaN (Not-a-Number), and
subnormal numbers. These special values are detailed in
Table~\ref{tab:fp8e5m2-special}.
\begin{table}\centering
    \begin{tabular}{|l|r||l|r|} \hline
        $\infty$ / Inf & $0\,11111\,00_2$ & $-\infty$/-Inf & $1\,11111\,00$
        \\\hline
        NaN & $\{0,1\}\,11111\,\{01, 10, 11\}$ & Zero & $\{0,1\} 00000 00$
        \\\hline
        Max normal & $\{0,1\}\,11110\,11 \equiv \pm 57\,344$ &
        Min normal & $\{0,1\}\,00001\,00 \equiv \pm 2^{-14}$ \\\hline
        Max subnormal & $\{0,1\}\,00000\,11 \equiv \pm 0.75\cdot 2^{-14}$ &
        Min subnormal & $\{0,1\}\,00000\,01 \equiv \pm 2^{-16}$ \\\hline
    \end{tabular}
    \caption{Special values for $E$ and $M$ in FP8\_E5M2}
    \label{tab:fp8e5m2-special}
\end{table}

The conversion from binary32 (or any other floating-point type that can be
expressed as shown in Equation~\eqref{eqn:fp-repr}) is performed as follows.
First, the exponent is checked to ensure it is within the allowed range. If the
unbiased exponent is greater than 15, the number is represented as infinity
($\text{Inf}$) in FP8\_E5M2; if it is less than -16, it is mapped to zero. For
exponents of -15 or -16, a subnormal number is generated. The mantissa of the
input number is truncated to its two most significant bits; however, to
implement proper round-to-even rounding, the subsequent three bits~--~the
\emph{guard}, \emph{round}, and \emph{sticky} bits~--~must also be considered.
Depending on these bits, the mantissa may be incremented by one, which may
consequently require an update to the exponent. Conversely, converting from
FP8\_E5M2 back to a higher-precision type is straightforward: the $S$, $E$, and
$M$ bits are simply placed into the corresponding positions of the target
precision type.

The basic arithmetic operations are implemented directly using the
representation defined in~\eqref{eqn:fp-repr}. For addition ($a+b$), we use
bit-arithmetic to align both mantissas (including extra bits) to the same
exponent. Subsequently, they are added or subtracted using integer arithmetic,
depending on the sign bits.

Afterwards, the mantissa is truncated according to the rounding rules and the
exponent is adjusted accordingly. These operations follow the standard
conventions for handling infinity ($\text{Inf}$) and not-a-number
($\text{NaN}$). Subtraction ($a-b$) is performed by inverting the sign of $b$
and then invoking the addition operation.

For normal numbers, multiplication $c = a \cdot b$ is performed as follows. We
first define
\[
\hat{M} = M + 2^m
\]
to include the implicit leading bit. The multiplication is then calculated as:
\begin{align}
    \label{eqn:fp8mult}
    a\cdot b &= \left(-1\right)^{S_a} \cdot 2^{E_a-B} \cdot (2^{-m} \cdot
    \hat{M}_a) \cdot \left(-1\right)^{S_b} \cdot 2^{E_b-B} \cdot (2^{-m} \cdot
    \hat{M}_b) \notag\\
    &= \left(-1\right)^{S_a \text{ xor } S_b} \cdot 2^{(E_a+E_b-2B)} \cdot
    (2^{-2m} \cdot \hat{M}_a \hat{M}_b) \notag \\
    &= \left(-1\right)^{S_c} \cdot 2^{E_c-B} \cdot (2^{-m} \cdot \hat{M}_c).
\end{align}
The result $c$ is then constructed from the quantities $S_c$, $E_c$, and
$\hat{M}_c$. Subnormal numbers are handled similarly, although they require
adjustments to the exponent and mantissa to fit the formulation
in~\eqref{eqn:fp8mult}. Operations on $S_c$, $E_c$, and $\hat{M}_c$ are
performed using either bit-wise operations or integer arithmetic.

In the case of division, the implementation follows the logic used
in~\eqref{eqn:fp8mult}, but the integer division is modified to avoid round-off
errors resulting from truncation. Specifically, the mantissa of the numerator
is shifted by $2^{2m}+3$ bits, where the three additional bits are required to
perform rounding using the guard, round, and sticky bits. Through these
methods, the basic arithmetic operations are implemented.

Since an $\text{FP8\_E5M2}$ value can only take one of 256 possible values,
mathematical functions such as $\sin(x)$ and $\sqrt{x}$ are implemented using
look-up tables. Consequently, each look-up table for a function with a single
$\text{FP8\_E5M2}$ input requires only 256 bytes. Certain functions, such as
$\operatorname{hypot}$, take two $\text{FP8\_E5M2}$ inputs and thus require 64
KiB of memory. In total, the look-up tables for all mathematical functions
described in Section~\ref{sec:user-defined-types} occupy less than one
megabyte. Comparison operations are also implemented using the bit
representation, as the absolute values of the $\text{FP8\_E5M2}$ type maintain
the same relative order as integers from 0 to 127. Thus, after accounting for
the sign, infinities, and NaN values, a simple integer comparison is
sufficient. The remaining functions from Section~\ref{sec:user-defined-types}
can be implemented based on these primitives.

The implementation is not restricted to half-precision or fp8 formats.
Similarly, any floating-point type supported by the C compiler can be mapped to
Fortran. Furthermore, Subsection~\ref{subsec:nocompiler} demonstrates how this
can be achieved using bit representations. Should future support for mapping
data directly to hardware floating-point units become available, the abstract
type introduced in Section~\ref{sec:user-defined-types} can be extended to
incorporate additional information. This would enable the definition of decimal
types, interval types, and other more complex objects. Additionally, by
grouping two elements of the same type within the abstract type, this approach
can be extended to support complex arithmetic, having hardware compatibility in
mind.

\section{Linear Algebra Operations}
\label{sec:blas}
Because many scientific computing applications rely on
BLAS~\cite{LawHKK79,DonDDH90,DonDHH88}, Low Precision Fortran incorporates
equivalent BLAS functionality. Conventionally, BLAS functions use a prefix
character to specify the floating-point data type. However, a single-character
prefix seems to be insufficient to identify various formats, such as FP8\_E5M2
and FP8\_E4M3, in a human-readable way, even though prefixes like \texttt{h}
(IEEE binary16) and \texttt{b} (BFloat16) are intuitive. Consequently, we
opted to eliminate these prefixes and encapsulate the BLAS functions within an
\texttt{interface} block,
allowing the compiler to resolve the correct implementation based on the
function's calling signature, as previously used for overloaded operators in
Section~\ref{sec:user-defined-types}. Since the user-defined data types behave
as standard floating-point types, existing single-precision BLAS routines can
be refactored for use with these types. To avoid duplicating this process for
each supported format, we employ a placeholder \texttt{DT}, which is mapped to
the target data type via a preprocessor definition during compilation.
Figure~\ref{fig:axpy} illustrates this method using a concise implementation of
the \emph{axpy} ($y\leftarrow \alpha x + y$) routine.
\begin{figure}
    \begin{subfigure}[t]{0.48\textwidth}
\begin{lstlisting}[language=fortran,basicstyle=\footnotesize\ttfamily,
    morekeywords={real32}]
subroutine saxpy(n, sa, sx, incx, sy, incy)
  integer, intent(in) :: n, incx, incy
  real(real32), intent(in) :: sa
  real(real32), intent(in) :: sx(*)
  real(real32), intent(inout) :: sy(*)
  integer :: i, ix, iy

  ix = 1
  iy = 1
  if (incx .lt. 0) ix = (-n+1)*incx + 1
  if (incy .lt. 0) iy = (-n+1)*incy + 1
  do i = 1,n
    sy(iy) = sy(iy) + sa*sx(ix)
    ix = ix + incx
    iy = iy + incy
  end do
end subroutine
\end{lstlisting}
        \caption{Single Precision}
        \label{fig:axpy:single}
    \end{subfigure}
    \hfill
    \begin{subfigure}[t]{0.48\textwidth}
\begin{lstlisting}[language=fortran,basicstyle=\footnotesize\ttfamily,
morekeywords={DT}]
subroutine axpy(n, sa, sx, incx, sy, incy)
  integer, intent(in) :: n, incx, incy
  type(DT), intent(in) :: sa
  type(DT), intent(in) :: sx(*)
  type(DT), intent(inout) :: sy(*)
  integer :: i, ix, iy

  ix = 1
  iy = 1
  if (incx .lt. 0) ix = (-n+1)*incx + 1
  if (incy .lt. 0) iy = (-n+1)*incy + 1
  do i = 1,n
    sy(iy) = sy(iy) + sa*sx(ix)
    ix = ix + incx
    iy = iy + incy
  end do
end subroutine
\end{lstlisting}
        \caption{Low-Precision with Data Type Placeholder}
        \label{fig:axpy:low}
    \end{subfigure}
    \caption{Concise \emph{axpy} implementation}
    \label{fig:axpy}
    \vskip-1.5ex
\end{figure}

As stated in the initial goals of the project, the number of modifications
required to switch between a compiler-supported floating-point type and a
user-defined one should be $\mathcal{O}(1)$. The \emph{axpy} example in
Figure~\ref{fig:axpy} demonstrates that this holds true for BLAS
implementations as well. Specifically, the only necessary changes are to the
data types of the \texttt{sa}, \texttt{sx}, and \texttt{sy} arguments.

On the one hand, the refactoring makes existing routines available in low
precision in an efficient manner. While the runtime overhead is not negligible,
it remains acceptable for proofs of concept and numerical evaluations. To
reduce this overhead, we employ the strategies presented in
Section~\ref{sec:implementation}. If the C compiler does not natively support
the low-precision data type, we use the refactored Fortran implementation based
on user-defined data types; this is specifically the case for the FP8\_E5M2 and
FP8\_E4M3 floating-point types. Conversely, if the C compiler provides native
support, such as for IEEE binary16 or BFloat16, we leverage the property
established in Section~\ref{sec:implementation} and
Equation~\eqref{eqn:vector}, which allows for the direct transfer of arrays
from Fortran to C. We then translate the Fortran implementation of the BLAS
routines into C and replace the data types with their C equivalents. This
approach enables the compiler to generate optimized code without calling helper
functions for basic arithmetic operations (e.g., addition and multiplication).
Furthermore, since some BLAS implementations, such as Intel's oneMKL, AMD BLIS,
or OpenBLAS~\cite{openblasweb}, provide only a limited subset of routines,
most-likely the general matrix-matrix multiply, for
certain low-precision data types, we include wrappers to these implementations
at compile time if they are detected. This is possible because our data types
rely on a memory-level representation that is consistent with the hardware.

\subsection{Interoperability with \tlapack}
\label{subsec:tlapack}
The \tlapack{}~\cite{tlapack} project aims to provide a C++ template-based
rewrite of the LAPACK functionality. The primary objective is to implement
LAPACK in C++ using templates so that by specifying a concrete data type
\texttt{T} at compile time, the required functions are instantiated for the
desired type. Consequently, the data type \texttt{T} must fulfill the
requirements of a proper numerical type in C++, similar to those described in
Section~\ref{sec:user-defined-types}. By providing a C++ header that implements
IEEE binary16, BFloat16, FP8\_E5M2, and FP8\_E4M3, data types from Low
Precision Fortran can be utilized within \tlapack{}. Since the memory layout is
designed to ensure seamless array handling in both Fortran and C, this
compatibility also extends to C++. By employing column-major storage in
\tlapack{}, matrices using all low-precision types can be handled consistently
across both Fortran and \tlapack{}. The mapping of data types from Fortran to
C++ is shown in Table~\ref{tab:dt-map}. By defining a proper interface on the
Fortran side, \tlapack{} code can be used seamlessly. Furthermore, the C++ data
type definitions can be used independently of the Fortran implementation to
provide OCP FP8\_E4M3 and FP8\_E5M2 support in C++.
\begin{table}
    \centering
    \begin{tabular}{|l|l|l|}\hline
      Data Type  & Fortran & C++ \\\hline
      IEEE binary 16 & \texttt{type(fp16)} & \texttt{lpf\_fp16\_t} \\
      BFloat16 & \texttt{type(bf16)} & \texttt{lpf\_bf16\_t} \\
      OCP FP8\_E5M2 & \texttt{type(fp8\_e5m2)} & \texttt{lpf\_fp8\_e5m4\_t}
      \\
      OCP FP8\_E4M3 & \texttt{type(fp8\_e4m3)} & \texttt{lpf\_fp8\_e4m3\_t}
      \\\hline
    \end{tabular}
    \caption{Data type mapping between Fortran and C++}
    \label{tab:dt-map}
\end{table}

\section{Usage and Example}
\label{sec:usage}
This section provides a concise example how to use the Low Precision
Fortran~\cite{lpf} library.
Both the example code and the library prioritize the ease of
implementing low-precision data types over peak performance, enabling users to
experiment without the need to master complex frameworks. The library requires
recent C, C++, and Fortran compilers that are ABI-compatible. These
requirements are met by GCC version 9 and later, LLVM clang/flang version 20
and later, and the Intel Compiler from the 2025 release. Although the library
is compatible with all CPUs, the C-compiler-based emulation of IEEE binary16
and BFloat16 requires specific modern CPU features; on x86\_64 architectures,
these features are available on nearly all CPUs released after 2016. To utilize
low-precision data types and the BLAS support in Fortran, include the
following \texttt{use}
statement:
\begin{lstlisting}[language=fortran]
    use lpf_XX
    use lpf_blas_XX
\end{lstlisting}
where \texttt{XX} corresponds to one of the type names listed in
Table~\ref{tab:dt-map}. Additionally, the code must be linked against the
\texttt{lpf} or \texttt{lpf\_blas} library, as appropriate. All types can be
used simultaneously without conflicts. Since the BLAS module does not have
data type prefixes in subroutine names, the code remains
independent of the selected data type. This allows the concrete data type to be
specified via a preprocessor definition.

\subsection{Implementing an Iterative Solver}
\begin{lstlisting}[language=fortran,
 float=t,
 caption={Conjugate Gradient in BFloat16},
 label={ex:cg},
 basicstyle=\footnotesize\ttfamily,
 commentstyle=\footnotesize\itshape,
 breaklines=true,belowskip=-3ex]
subroutine cg_bf16(n, A, lda, x, maxit, tol)
  use lpf_bf16, lpf_blas_bf16

  integer, intent(in) :: n, lda, maxit
  type(bf16), intent(in) :: A(lda, n), tol
  type(bf16), intent(inout) :: x(n)

  type(bf16), allocatable :: r(:), p(:), Ap(:), b(:)
  type(bf16) :: rho, rho_new, alpha, beta, one, zero
  integer :: k

  one = bf16(1.0)
  zero = bf16(0.0)
  allocate(b(n), r(n), p(n), Ap(n))

  b = x
  x = zero
  r = b
  call gemv('N', n, n, -one, A, lda, x, 1, one, r, 1)
  p = r
  rho = dot(n, r, 1, r, 1)

  do k = 1, maxit
    call gemv('N', n, n, one, A, lda, p, 1, zero, Ap, 1)          ! Ap = A * p
    alpha = rho / dot(n, p, 1, Ap, 1)                ! alpha = rho / (p' * Ap)
    x = x + alpha * p
    call axpy(n, -alpha, Ap, 1, r, 1)                     ! r = r - alpha * Ap
    rho_new = dot(n, r, 1, r, 1)                            ! rho_new = r' * r

    write(*,'(A, I5, A, DT"E"(10,5),A, DT"E"(10,5))') "IT = ", k,  &
          &  " RHO_NEW = ", sqrt(rho_new,)  " TOL = ", tol
    if (sqrt(rho_new) .lt. tol) exit

    beta = rho_new / rho
    p = r + beta * p
    rho = rho_new
  end do
  deallocate(b, r, p, Ap)
end subroutine
\end{lstlisting}
In addition to the AXPY implementation shown in Figure~\ref{fig:axpy}, we now
present a more complex example. The objective is to solve a linear system $Ax =
b$, where $A$ is a symmetric positive-definite matrix, using the Conjugate
Gradient algorithm~\cite{Saa03a}. Listing~\ref{ex:cg} demonstrates a
straightforward implementation that combines Fortran array notation and BLAS
calls in BFloat16. By modifying the data type definitions, the example can be
used for IEEE binary16, FP8\_E5M2, and FP8\_E4M3 without further modification.

To provide a well-defined example, we set $A$ and $b$ as:
\[
A = \begin{bmatrix}
    2 & -1 & & & \\
    -1 & 2 & -1 & & \\
    & \ddots & \ddots & \ddots & \\
    & & -1 & 2 & -1\\
    & & & -1 & 2
\end{bmatrix} \text{ and } b = \begin{bmatrix} 1  \\ 0 \\ \vdots \\ 0 \\
    1\end{bmatrix},
\]
which yields the results shown in Table~\ref{tab:ex-cg} for $n=20$ and
$\texttt{tol} = 10^{-3}$.
\begin{table}\centering
    \begin{tabular}{|l|r|r|}\hline
        Data Type & Iterations & $\dfrac{||Ax-b||_2}{||A||_F + ||b||_2}$
        \\\hline
        IEEE Binary 16 & 11 & $2.55 \cdot 10^{-4}$\\
        BFloat 16 & 15 & $3.18\cdot 10^{-3}$\\
        FP8\_E5M2 & 14 & $2.54\cdot 10^{-2}$ \\
        FP8\_E4M3 & 66 & $1.22\cdot 10^{-1}$\\\hline
    \end{tabular}
    \caption{Result of the Conjugate Gradient Example with different data types}
    \label{tab:ex-cg}
\end{table}
As expected, the accuracy of the computed solution decreases from IEEE binary16
to the FP8 types. Again, future adjustments to support if the data type is
supported by the Fortran compiler, restricts to the adjustment of the
\texttt{type(bf16)} declarations and the format specifiers in the
\texttt{write} statement.

\section{Conclusion}
The Low Precision Fortran software library enables the use of common
low-precision data types. Unlike other libraries, such as GNU
MPFR~\cite{FouHLetal07}, this work focuses on a storage scheme that allows the
internal representation of values to be used in both software and hardware.
This design also enables the library to serve as the CPU-side component of
GPU-accelerated code, where low-precision data types are often supported in
hardware. The examples provided in this work demonstrate that low-precision
floating-point arithmetic can be used in Fortran without the need for
specialized frameworks. By providing BLAS functionality and ensuring
compatibility with C++, the library allows users to develop and evaluate the
behavior of low-precision floating-point types in projects that extend beyond
trivial examples. Once native support for these low-precision types becomes
available in Fortran compilers, the implemented code can be easily refactored.

Although the primary focus was on low-precision types with hardware
representations, the proposed concept is not restricted to them. Provided that
all functions and operators mentioned in Section~\ref{sec:user-defined-types}
are implemented, this framework can be applied to arbitrary floating-point
types. For instance, if operations are implemented using GNU
MPFR~\cite{FouHLetal07}, the same approach can be employed to implement
ultra-high-precision arithmetic, which then integrates seamlessly into Fortran.

Future work will involve extending the library to include the most commonly
used LAPACK routines. We also encourage BLAS implementers to provide more
than the optimized GEMM routines for types with precision lower than IEEE
binary32. Furthermore, we plan to support complex numbers based on the provided
low-precision types as mentioned in Section~\ref{sec:implementation}.

\section*{Acknowledgments}%
\addcontentsline{toc}{section}{Acknowledgments}
    We thank Eda Oktay, whose inquiry regarding the possibility of performing
    low-precision arithmetic in Fortran codes served as the catalyst for this
    project.


\addcontentsline{toc}{section}{References}
\bibliographystyle{plainurl}
\bibliography{paper_final}

\end{document}